\documentclass[12pt]{iopart}
\usepackage{graphicx}
\usepackage{iopams}
\usepackage{color,graphicx}

\begin{document}

\title[Quantum criticality in the iron pnictides and chalcogenides]
{\topical{Quantum criticality in the iron pnictides and chalcogenides}}

\author{Elihu Abrahams$^{1}$ and Qimiao Si$^{2}$}
\address{$^1$ Department of Physics and Astronomy, University of California Los Angeles, Los Angeles, California 90095, USA}
\address{$^{2}$ Department of Physics and Astronomy, Rice University, Houston, Texas 77005, USA}
\ead{abrahams@physics.ucla.edu, qmsi@rice.edu}

\begin{abstract}
Superconductivity in the iron pnictides and chalcogenides arises at the border of antiferromagnetism, which raises the question of the role of quantum criticality. In this topical review,
we describe the theoretical work that led to the prediction for a magnetic quantum critical point
arising out of a competition between electronic localization and itinerancy, and the proposal
for accessing it by using isoelectronic P substitution for As in the undoped iron pnictides. We go on to compile the emerging experimental evidence in support of the existence of such a quantum critical point in
isoelectronically-tuned iron pnictides. We close by discussing the implications of these results for the
physics
of the iron pnictides and chalcogenides.

\end{abstract}
\pacs{}
\submitto{\JPCM}
\maketitle

\section{Introduction}
In recent years, stimulated by the discovery of new materials with unusual properties, the physics of ``quantum criticality" has become a frontier issue in condensed matter physics. In many materials, several ground states are often in competition and it is usually possible to tune from one ground state to another by adjusting some control parameter.   Such a transition from one zero temperature phase to another is called a quantum phase transition;
when the transition is continuous,
it defines a quantum critical point \cite{subir}. Typical control parameters include magnetic field, pressure, electron density, and chemical composition. In contrast to ordinary
thermally driven
phase transitions, where the control parameter is temperature,
dynamical fluctuations of order parameters are on equal footing with the usual spatial fluctuations.
The quantum critical scaling that describes the neighborhood of quantum critical points
has been widely discussed
starting from the very early work of John Hertz \cite{hertz}.
A recent review (in the context of Fermi-liquid instabilities) \cite{hvl} summarizes
important aspects of quantum criticality.
Just as spatial fluctuations are characterized by a correlation length $\xi$ that diverges at a phase transition governed by a control parameter $x$,
\numparts
\begin{equation}
\xi \propto (x-x_c)^\nu,
\end{equation}
dynamical (temporal) fluctuations are characterized by a diverging correlation time
\begin{equation}
\tau \propto (x- x_c)^{z\nu}.
\end{equation}
\endnumparts
Here, $z$ is the
dynamic
exponent of the quantum phase transition.

Examples of well-studied quantum phase transitions are the zero temperature transitions in quantum Hall effect systems \cite{qhe}, in heavy fermion compounds \cite{hf}, in the cuprate high-temperature superconductors \cite{hitc}, and, in disordered metals, the metal-insulator transition \cite{mit}.
Theoretical work has advanced the notion that
Landau's paradigm of order-parameter fluctuations, which works so well
for thermally-driven phase transitions, can fail for quantum critical point \cite{si01,senthil04}.  Consequently,  a unified theory of quantum criticality does not yet exist, and many
 fundamental questions are
 open, for example:
What collective excitations are
generally important at and near a quantum critical point?
Are there universality classes?
What are the scaling laws and their exponents?
Under the circumstances, the investigation of different classes of quantum critical phenomena in new materials can be an avenue to further understanding.

The discovery \cite{Kamihara_FeAs,Zhao_Sm1111_CPL08} of a number of iron-based compounds that exhibit superconductivity at relatively high temperatures (compared to ``conventional" superconductors) has generated a large body of experimental and theoretical research \cite{ishida,greene} and because of the presence of competing phases in these compounds we have a new window into quantum critical phenomena.

\begin{figure}[h]
\centering
\includegraphics[totalheight=0.38\textheight, viewport= 90 150 800 580,clip] {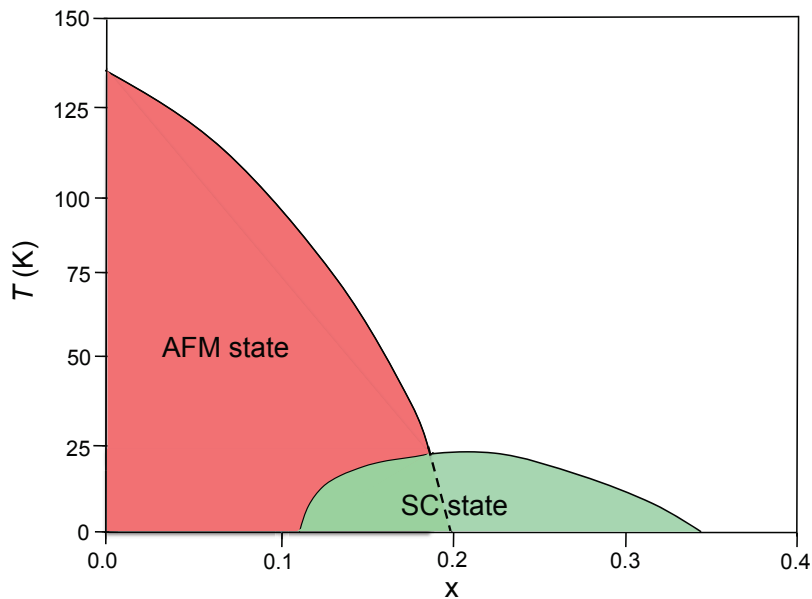}
\caption
{{\bf Phase diagram of Ba(Fe$_{1-x}$Co$_x$)$_2$As$_2$}. Adapted from Wang {\it et al} \cite{xfwang}
and Chu {\it et al} \cite{jhchu};
the antiferromagnetic (AFM) state is also structurally distorted.
 In this case, the quantum critical point at $x\approx 0.2$ is obscured by the onset of superconductivity
 with a maximum transition temperature of $\approx 25$ K.
 The dashed line is the expected continuation of the magnetic transition
line in the absence of superconductivity.
 }
\label{CoDope}
\end{figure}

The phase diagrams of the iron pnictides rather universally exhibit an antiferromagnetic ground state for the undoped (parent) compounds, {\it e.g.} LaFeAsO (1111), BaFe$_2$As$_2$ (122). Upon substitution of F for O in the 1111 case and K for Ba or Co for Fe in the 122 case, the ordered moment gradually disappears as a critical doping level is approached. However superconductivity intervenes in this doping region and magnetic quantum criticality is not so evident. In figure \ref{CoDope}, the behavior of Co doped 122 is
shown \cite{xfwang,jhchu}, which illustrates this point.
Therefore it would be advantageous to seek an avenue for the exploration of the magnetic quantum critical point that is not obscured by the influence of other phases.

The influence of electron-electron interactions, {\it i.e.} the consequences of correlation effects, plays a central role in our discussion and we begin with remarks that review that issue. Then we discuss some consequences of proximity to a Mott transition. This is followed by considerations of a magnetic quantum criticality that is tuned by electron-electron interaction strength rather than by the usual parameters mentioned above. We give examples in the iron-based compounds and  conclude with a summary and outlook.

\section{Iron pnictides as bad metals}

The materials under discussion in this review are paramagnetic metals at temperatures greater than about 200 K and usually become antiferromagnets at lower $T$. When doped away from their stoichiometric compositions the magnetism disappears and superconductivity often arises. A recent review \cite{ishida} summarizes the physical properties of the various compounds. Soon after the discovery \cite{Kamihara_FeAs} of superconductivity in the iron arsenide compound, we pointed out \cite{sieaprl} that electron-electron interaction has a significant effect on these properties. Subsequently, we emphasized and reviewed comprehensively \cite{njp} the evidence for the importance of correlation effects. In this section, we briefly summarize those considerations.

What are diagnostics to evaluate the strength of electron-electron correlation in a metal? One is the magnitude of the room temperature resistivity. Electron-phonon scattering rarely leads to a room temperature mean free path $\ell$ that is of order the lattice spacing. Therefore, a small value of $k_F \ell$ can be an indicator of strong electron-electron interaction. In the pnictides, there are five bands that cross the Fermi surface. For a rough estimate, let us assume that the corresponding Fermi pockets all have the same value of $k_F \ell$. The in-plane conductivity \cite{sieaprl} is the sum of contributions of each pocket, say $p$ of them:
\[
\sigma = p(e^2/h)k_F \ell /c,
\]
where $c$ is the $c$-axis lattice spacing. A typical value for the single-crystal in-plane conductivity
at room temperature is 2500 (Ohm-cm)$^{-1}$ and $c$
is about 6.5  \AA \; for 122 iron arsenides and
about 8.5 \AA \;  for 1111 iron arsenides.
We get $k_F\ell \approx 5/p$, which gives $k_F \ell$
of order unity for each pocket. Hence, the pnictides may properly be described as ``bad metals" as a consequence of strong electron-electron interactions.

Another indication of the importance of correlation comes from a diminished strength of the Drude peak of the optical conductivity.  A particularly comprehensive discussion of this maybe found in a paper of Qazilbash  {\it et al} \cite{qaz}
(see also, reference~\cite{degiorgi,ferber}),
in which the pnictides are compared to well-known strongly and moderately correlated materials such as the cuprate superconductors and V$_2$O$_3$.
The reduction of the Drude weight provides a measure of the distance of the bad-metal system from the Mott transition point. By this measure, BaFe$_2$As$_2$, for instance,
which has only about 30\% of the electronic excitations residing in the coherent part  of the optical spectrum \cite{qaz}, has interactions that are about 70\% of the Mott-transition value.

The reduction of Drude weight is accompanied by unusual features in the temperature-induced transfer of spectral weight. It is observed \cite{hu,boris,yang} that as the temperature is lowered, there is a transfer of optical spectral weight from low to high (on the order of eV) energies.
This  is  expected as a precursor to the formation of Hubbard bands \cite{roz} and Mott-Hubbard localization.

\section{Mott transition}

\begin{figure}[h]
\centering
\includegraphics[totalheight=0.3\textheight, viewport=-50 120 800 470,clip]
{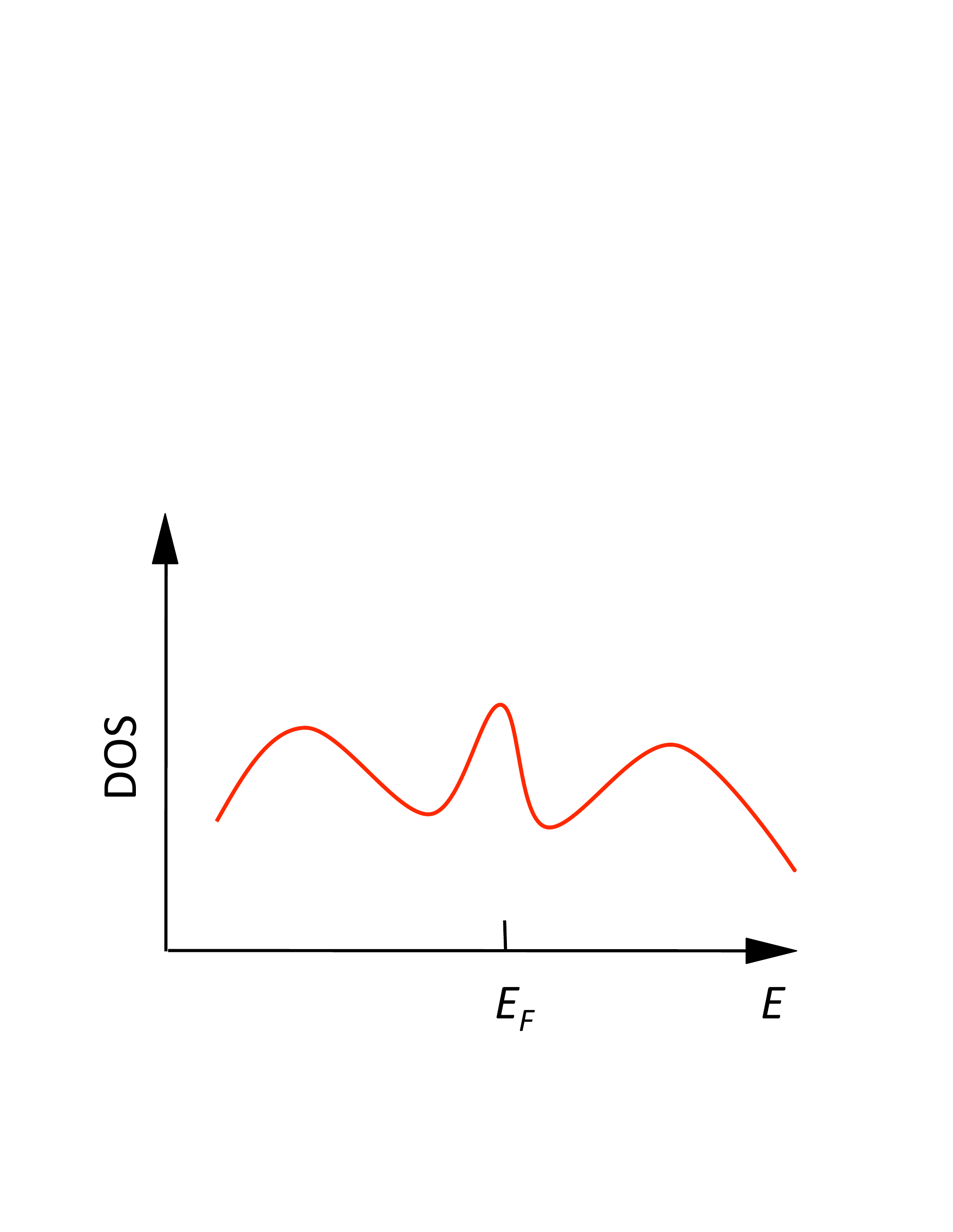}
\caption
{{\bf Single-electron spectral function as the sum of
coherent and incoherent parts.}
The single-electron density
of states (DOS) is plotted against energy ($E$); $E_F$ is the
Fermi energy.
Each peak may
contain additional structure
due to the multi-orbital nature of the iron
pnictides.
}
\label{dos}
\end{figure}

If the electron-electron interaction is sufficiently strong, it is expected that the ground state of the system will be a Mott insulator, characterized by upper and lower Hubbard bands separated by an energy gap - the so-called Mott gap, of order the interaction strength
minus the bandwidth.  There are many examples, notably the undoped cuprate superconductors and vanadium sesquioxide. The ratio of the interaction strength to the bandwidth, {\it i.e.} the ratio of potential to kinetic energy, as usual, measures the proximity to the Mott localization transition. On the localized side of the transition, the low lying excitations are of the degrees of freedom associated with localized spins, whose spectral weight
comes from
the lower Hubbard band. We have argued \cite{sieaprl,njp} that the iron pnictides are on the metallic side of the Mott transition, but very close to it. In this picture, the electron spectral function has a three peak structure consisting of the precursors of the Hubbard bands away from the Fermi level $E_F$ and a quasiparticle peak at $E_F$ that is responsible for the metallic behavior, as depicted in figure \ref{dos}. It is then natural to ascribe \cite{sieaprl} the magnetism of the undoped metallic iron-based compounds to the behavior of the local moments associated with the developing lower Hubbard band.

The situation depicted in figure \ref{dos} is a simplified sketch of what has been determined from electron structure studies by a Rutgers University group that has taken correlation effects into account systematically \cite{DMFT}.
In that work, in analyses of the optical conductivity, the authors find a strongly reduced Drude peak and substantial incoherent spectral weight smeared over energies of the order of an electron volt
(see also, reference~\cite{laad}).
The absence of well-defined features associated with incipient localized 3$d$ states is due to their crystal field splittings as well as broad features from the As 4$p$ states. Their analysis also indicates that although the system is metallic, the sublattice magnetization with reduced moment resides on the iron atoms.

The standard theoretical description for Mott transition occurs in one-band models with an odd number of electrons per site. The parent iron pnictides and chalcogenides contain multiple orbitals that are partially occupied by an even number of (six) 3$d$ electrons per Fe site, and any insulating state could in principle simply be a band insulator. Multi-orbital models for the iron pnictides have been studied in Ref.~\cite{yu} with the conclusion that a Mott transition robustly exists as a function of $U/t$.

The description of the properties near the Mott transition suggests that by tuning the ratio of potential to kinetic energy (``PE/KE") one might be able to access two quantum phase transitions.  One is the Mott metal-insulator transition itself; the other is a magnetic transition from antiferromagnet to paramagnet on the metallic side as one moves away from the Mott state. In this way, we interpret the observed magnetism in the undoped arsenides as due to quasilocalized spins belonging to the precursor of the lower Hubbard band, as shown in figure \ref{dos}. In this paper we begin with the magnetic transition and discuss the Mott transition near the end in connection with recent experiments on iron oxychalcogenides.

The principal method of tuning PE/KE that we consider here is by isoelectronic chemical substitution that affects the lattice parameters. The advantage over charge doping ({\it e.g.} F for O, K for Ba, Co for Fe) is that neutral impurities create much less disorder than charged ones. So much so that deHaas van Alphen experiments succeed over a substantial range of P doping for As (see section 5.1) and NMR lines remain narrow \cite{NakaiPRB}. Specifically, upon substitution of
smaller phosphorous for
larger arsenic in an iron arsenide, the lattice constant decreases; therefore the electron bandwidth, the KE in the tuning ratio, increases. At the same time, the intra-atomic interactions responsible for correlation effects remain essentially constant. Thus, one may pass from the antiferromagnetic state of the arsenide to the putative paramagnetic state of the phosphide via a quantum critical point at a critical concentration of phosphorous. This was the prediction made in \cite{pnas} and subsequently confirmed experimentally \cite{exp}. The evidence is reviewed in section 5.

\section{Theory of the magnetic quantum critical point}

The antiferromagnetism in a typical
iron pnictide
is characterized by ${\bf Q}=(0,\pi)$ or $(\pi,0)$ order on a square lattice \cite{exp}. That is, in one direction there is antiferromagnetic order and in the orthogonal direction, ferromagnetic order. For the present, we ignore the orthorhombic distortion that accompanies the magnetism. Consistent with the discussion of the previous section, in which the magnetism arises from the localized spins in the precursor Hubbard band, the exchange coupling
naturally involves first ($J_1$) and second neighbor ($J_2$) spins with $J_2>J_1/2$
\cite{sieaprl}.
The magnetic part of the Hamiltonian is then \cite{foot1}
\begin{equation}
{\cal H}_{mag} = \sum_{nn}J_1 S_i\cdot S_j + \sum_{nnn}J_2 S_i\cdot S_j.
\label{ham}
\end{equation}
This model, with the prescribed ratio between $J_2$ and $J_1$, naturally generates the collinear order \cite{chris,ccl}, and was proposed to underlie the magnetic order in the parent iron
pnictides \cite{sieaprl}.

It is worth underscoring the preconditions that make this spin Hamiltonian the starting point of the description for the spin properties of the parent iron pnictides \cite{sieaprl,pnas}. The considerations in the previous two sections imply that the spin spectral weight will primarily come from electronic states away from the Fermi energy. We use a parameter $w$ to measure the coherent part of the electronic spectral weight near the Fermi energy ({\it cf.} figure \ref{dos}). $w$ is relatively small when the electronic correlation is close to the threshold value for the Mott transition. It is also small because the Fermi surfaces of the parent iron pnictides (and chalcogenides) comprise small electron and hole pockets. The weight $w$ can therefore be used as the perturbation parameter, in place of $U/t$ that is order unity. The spin Hamiltonian (\ref{ham}) arises in the zero-th order of $w$. The effect of higher order terms on the spin properties is treated below; {\it cf.} eq
 uation (\ref{gl}).

\begin{figure}[h]
\centerline{
\includegraphics[width=.73\linewidth,
viewport=103 -8 670 520, clip]
{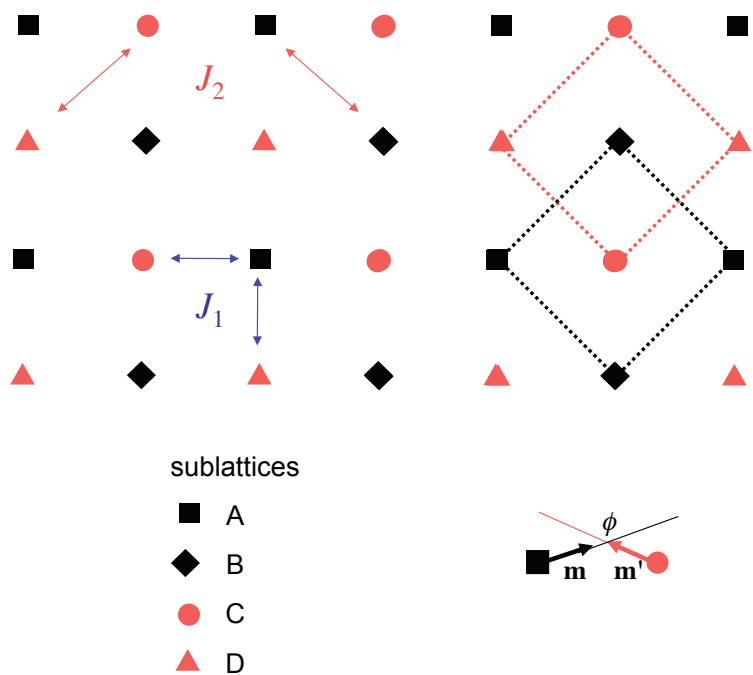}}
\vskip -.3in
\caption
{{\bf Classical ground state of the $J_1-J_2$ model.}
Two interpenetrating N{\'e}el square lattices with staggered magnetizations ${\bf m}$ and ${\bf m}'$, where $\phi$ is the angle between them.}
\label{mag}
\end{figure}

As discussed in \cite{ccl}, when an antiferromagnetic $J_2$ exceeds $J_1/2$, the classical ground state consists of two interpenetrating N{\' e}el lattices (the dotted squares in figure \ref{mag}) with independent staggered magnetizations (N{\'e}el vectors) ${\bf m}$ and ${\bf m}'$. Thus, the order is characterized by four sublattices imbedded in a square lattice, as shown in figure \ref{mag}. Here, sublattices $A$ and $B$ are antiferromagnetically coupled with staggered magnetization ${\bf m}$ as are $C$ and $D$ with ${\bf m}'$. Within the $J_1-J_2$ model, the energy is independent of the angle $\phi$ between ${\bf m}$ and ${\bf m}'$. The ultimate collinear order in the quantum system is achieved by the phenomenon called ``order from disorder" \cite{villain}. Here, the ``disorder" refers to thermal or quantum fluctuations \cite{chris} that break the degeneracy in $\phi$ and favor the collinear arrangement $\phi=0$ or $\pi$. Thus ${\bf m}\cdot {\bf m}' = \pm 1$ (or equivalent
 ly $\sigma$, where ${\bf m}' = \sigma\,{\bf m}$ \cite{Xu:08}) becomes an Ising order parameter and we shall include it in the Ginzburg-Landau theory that we develop below. The identification of an Ising transition in the frustrated $J_1-J_2$ model was first made in \cite{ccl} and that formulation was applied to the
pnictides in \cite{Xu:08,Fang:08}.

The Ising order breaks the $C_4$ symmetry and naturally leads to a tetragonal-to-orthorhombic distortion in the presence of a coupling between the Ising order parameter ${\bf m} \cdot {\bf m'}$ and
a structural degree of freedom. The structural distortion, perhaps in combination with orbital ordering among the Fe $d_{xz}$ and $d_{yz}$ states \cite{orb1,orb2}, can give rise to  a planar anisotropy in the value of $J_1$, {\it i.e.} $J_{1a}\neq J_{1b}$ \cite{orb1}, as has been suggested by a spin-wave analysis of the inelastic neutron-scattering spectrum of CaFe$_2$As$_2$ in its magnetically ordered phase \cite{zhao}.

\begin{figure}[h]
\centering
\includegraphics[totalheight=0.36\textheight, viewport=-80 120 750 570,clip]
{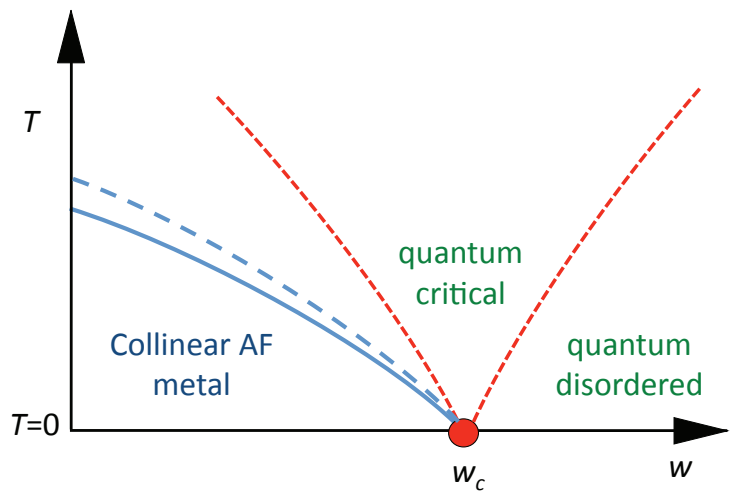}
\caption
{{\bf Pnictide phase diagram near a magnetic quantum critical point.}
The red dot denotes the quantum critical point determined by the critical tuning parameter $w_c$. The blue line is the
line of the thermally-driven
antiferromagnetic transition
and the dashed line is a structural transition.
At non-zero temperatures, the two transitions can
occur either coincidentally, or be separated. At zero temperature,  they coincide.
}
\label{phase}
\end{figure}

Equation (\ref{ham}) applies at the Mott transition point, where $w=0$. The effect of a non-zero $w$ can be treated order-by-order in terms of a Ginzburg-Landau action.
As we discussed in the previous section,
the quasiparticle weight $w$ at the Fermi energy seen in
figure \ref{dos} can be tuned by the strength of electronic correlation.
We propose that when $w$ is increased,
a quantum phase transition
out of the collinear
antiferromagnetic state occurs.
We consider the
lattice spacing
as a tuning parameter but the considerations here just depend on the spectral weight $w$ of the quasiparticle peak at the Fermi energy: the smaller the
lattice
parameter, the larger is the KE
(or, equivalently, the smaller is $U/t$)
and the larger is $w$ and we shall use it as a tuning parameter. With all these considerations, the quantum Ginzburg-Landau action at $T=0$ has the form \cite{njp,pnas}
\begin{eqnarray}\fl
  {\cal S}({\bf m},{\bf m}') =
 \int d\{\bf q\}
  \int d\{\omega\} [S_2({\bf q},\omega) + S_4 (\{{\bf q}\},\{\omega\}) + \dots], \nonumber \\
\fl
S_2({\bf q},\omega) = [r_0 + wA_{\bf Q} +c {\bf q} ^2 + \omega^2 + \gamma|\omega|][|{\bf m}({\bf q},\omega)|^2 + |{\bf m}'({\bf q},\omega)|^2] \nonumber \\
 +  v(q_x^2-q_y^2)[{\bf m}({\bf q},\omega)\cdot {\bf m'}({-\bf q},-\omega)], \nonumber \\
\fl S_4 (\{{\bf q}\},\{\omega\}) = u(|{\bf m}|^4 + |{\bf m}'|^4) + {\tilde u}|{\bf m}\cdot {\bf m}'|^2 + u'|{\bf m}|^2\; |{\bf m}'|\,^2.
\label{gl}
\end{eqnarray}
The ${\bf q}$-dependent terms in $S_2$ appear naturally from a gradient expansion of the energy represented in (\ref{ham}).  The notation $\{{\bf q}\},\{\omega\}$ denotes the set of four momenta and four frequencies that enter $S_4$. The
momentum and frequency integrals
 are understood to each contain a delta function that fixes the sum of the momenta and the sum of the frequencies at zero.
Several important comments on (\ref{gl}) are in order:

We take $r_0 < 0$, so that when $w=0$, the system is antiferromagnetically ordered. At a critical value  $w_c$, $r_0 + w_c A_{\bf Q} = 0$, which defines the quantum critical point separating the magnetic state from a non-magnetic one. A sketch of
the resulting
phase diagram is shown in figure \ref{phase}. The dashed red crossover lines border the region of quantum critical fluctuations and are defined by $\hbar/k_B T \sim \tau$, or $T \propto |w-w_c|^{\nu z}$, see (1b).

When $w$ is non-zero, the local moments are coupled to the itinerant quasiparticles; this leads to the damping term $\gamma|\omega|$ in $S_2$, where for low energies, $\gamma$ is O($w^0$) \cite{pnas}. This is valid as long as $|\omega|$ is less than the quasiparticle bandwidth $wW$. Here $W$ is the full bandwidth at $w=1$. Therefore, the critical theory has dynamical exponent $z=2$ \cite{foot2}, as is typical for an antiferromagnetic quantum phase transition.

As mentioned earlier, the non-zero temperature magnetic transition in the parent iron pnictides is accompanied or preceded by a tetragonal to orthorhombic structural distortion in which the square array of iron atoms becomes rectangular, where the longer $a$-axis is the direction of the eventual antiferromagnetic order. This can naturally be accommodated in the theory by including in (\ref{gl}) a coupling of the Ising order parameter ${\bf m}\cdot {\bf m}'$ to a structural degree of freedom.

The sign of the quartic coupling ${\tilde u}$ in (\ref{gl}) is very significant. It can be derived from the above $J_1-J_2$ model \cite{ccl}; it is negative, reflecting the physics of order from disorder inherent in the $J_1-J_2$ model \cite{ccl} and therefore leads to the Ising transition fixing ${\bf m}\cdot {\bf m}' = \pm 1$ as discussed above. It also gives rise to the collinear arrangement in the ordered state, in which the moments are antiferromagnetically ordered along the $a$-axis and ferromagnetically ordered along the $b$-axis of the planar Fe-Fe lattice.

In the traditional approach to quantum criticality \cite{hertz,millis} it is recognized that $d_{eff} = d+z$ is the effective dimensionality of critical fluctuations, where $d$ is the spatial dimension. In our $d=2,\,z=2$ case, the $d_{eff}$ is 4; therefore we are at the upper critical dimension at which there are logarithmic corrections to simple Gaussian critical behavior and some predictions can be made from scaling arguments \cite{hvl}.  Experimentally, the quantum critical behavior can be explored either by tuning $w$ through $w_c$ at very low temperature or by lowering the temperature with $w$ fixed at $w_c$. For example, the specific heat varies as
$-T \ln (w-w_c)$
and $-T\ln T$ respectively for the two cases. The resistivity is determined by the interplay of two scattering mechanisms: spin fluctuations and impurities. In the dirty limit, for $w=w_c$, the resistivity varies linearly with $T$ \cite{rosch}, $\rho - \rho_0 \propto T$, rather than quadratically as in a Fermi liquid. In this $d=z=2$ case, the uniform susceptibility is expected to vary as $\chi - \chi_0 \propto(-T\ln T)$ \cite{ioffe}.

\section{Accessing the quantum critical point by isoelectronic substitution}

We mentioned at the end of Section 3 that substitution of P for As decreases the lattice constant, thereby reducing the PE/KE ratio and increasing $w$. Table \ref{one} gives some representative numbers that illustrate the change in lattice parameter. The references are \cite{LaCeA} for the La and Ce arsenides ($x=1$), \cite{LaCeP} for the La and Ce phosphides, \cite{Ba} for Ba, and \cite{Eu} for the Eu compound.

\begin{table}[h]
\caption{\label{one}Lattice parameters for isoelectronic substitution.}
\begin{indented}
\lineup
\item\begin{tabular}{llllllc}
\br
Compound&$x$&$c(\AA)$&$a(\AA)$&$c/a$&$V(\AA^3)$&Reference\\
\mr
LaFe(As$_x$P$_{1-x}$)O&1&8.754&4.038&2.168&142.7&\cite{LaCeA}\\
&0&8.508&3.957&2.150&133.2&\cite{LaCeP}\\
\mr
CeFe(As$_x$P$_{1-x}$)O&1&8.656&4.000&2.164&138.5&\cite{LaCeA}\\
&0&8.328&3.919&2.125&127.9&\cite{LaCeP}\\
\mr
BaFe$_2$(As$_x$P$_{1-x})_2$&1&13.001&3.962&3.281&204.1&\cite{Ba}\\
&0&12.422&3.844&3.232&183.5&\cite{Ba}\\
\mr
EuFe$_2$(As$_x$P$_{1-x})_2$&1&12.136&3.910&3.104&185.8&\cite{Eu} \\
&0.7&11.831&3.889&3.042&178.9&\cite{Eu}\\
\br
\end{tabular}
\end{indented}
\end{table}
It can be seen that isoelectronic P substitution for As results in a substantial reduction in unit cell volume ($V$), which implies an increase in KE with little change in PE, which is primarily local. Thus as $x$ decreases, the tuning parameter $w$ increases. Since the As parent compounds exhibit antiferromagnetic order (small $w$ side of figure 3) and the P compounds do not,
we can place the two end materials on the two sides of $w_c$ along the tuning-parameter axis in the theoretical phase diagram, figure \ref{phase}. Consequently,
one expects at some value of $x=x_c$ (corresponding to $w_c$)
 to access a magnetic quantum critical point. Then the measurement of various properties at P concentration $x_c$ will give information on the quantum critical behavior, {\it e.g.} the critical exponents. Thus, the method of isoelectronic substitution offers a new route to quantum criticality, as distinct from
charge
carrier
doping or application of external pressure. At the same time, this approach is not necessarily free of the interference of other phases as discussed above in connection with figure \ref{CoDope}.

\subsection{122 compounds}
These are the iron pnictides of composition QFe$_2$Pn$_2$, where Q is usually Ba, but can also be a different valence 2 element {\it e.g.} Ca, Sr, and even Eu. To achieve superconductivity, the 122s are either electron doped by partial substitution of Fe by another transition metal ion, or hole doped by partial substitution of Q by an alkali metal ion. However, P doping for As in {\it e.g.} BaFe$_2$As$_2$ also causes the suppression of magnetism and the appearance of superconductivity. In spite of the interference of superconductivity at the critical P concentration $x_c$ for the disappearance of antiferromagnetism, some information has been obtained about transport and thermodynamic properties at $x=x_c$ in (Ba or Eu)Fe$_2$(As$_{1-x}$P$_x$)$_2$.

In EuFe$_2$(As$_{1-x}$P$_x$)$_2$ with $x=0.3$ \cite{Eu}, no trace of the AFM order remains and not only is there an apparent superconducting transition at $T=26$ K, but the Eu moments appear to order ferromagnetically at $T=20$ K. Nevertheless, at higher temperatures, the resistivity is linear, $\rho = \rho_0 + a T$, in temperature from about 90 K up to 300 K. However, the specific heat does not show signs of the $T\ln T$ behavior expected for the $d=2, z=2$ described above in section 4. While the appearance of the two phases that are induced by P-doping is interesting in itself, it obscures the magnetic quantum criticality associated with the suppression of the antiferromagnetic state.

A group at Zhejiang University  has studied the BaFe$_2$(As$_{1-x}$P$_x$)$_2$ series \cite{jiang}. The experimentally determined phase diagram is very similar to that shown in figure \ref{CoDope}. The principal difference is that the scale of doping should be rescaled by a factor 1.6. The fits to a resistivity of the form $\rho = \rho_0 + a T^{\alpha}$ give $\alpha$ near 2 (as in a Fermi liquid) for concentrations $x$ far from $x_c$ but near the critical concentration $x_c = 0.32$, a $T$-linear resistivity was found up to room temperature: $\alpha \approx 1$ for $0.3 < x< 0.45$.

The $T$-linear behavior of the resistivity in the critical concentration region of BaFe$_2$(As$_{1-x}$P$_x$)$_2$ was subsequently confirmed by a group from Kyoto University \cite{kasahara}. In fact, as emphasized in \cite{kasahara}, the region of the phase diagram in which the resistivity is linear in $T$ bears a striking resemblance to the quantum critical region shown in figure \ref{phase}, in spite of the presence of superconductivity at the critical concentration.

\begin{figure}[h]
\centering
\includegraphics[totalheight=.28\textheight, angle=0, viewport=5 230 650 570,clip]
{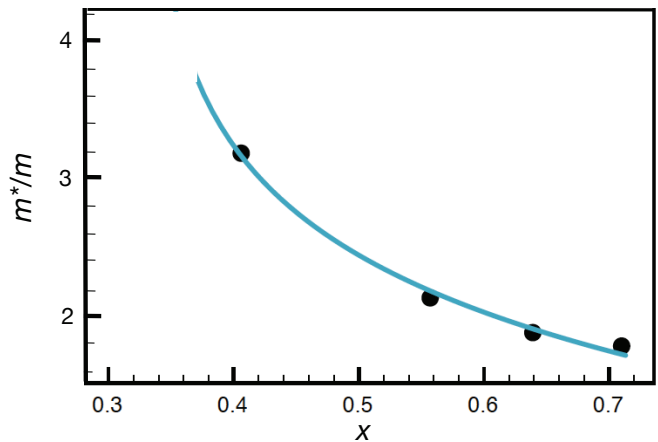}
\caption
{{\bf Effective mass of BaFe$_2$(As$_{1-x}$P$_x$)$_2$.}
The experimental data (black dots) is taken from figure 3b of reference \cite{shish}. A fit $m^*/m = 0.88-0.89\ln (x-0.33)$ is shown by the solid line.}
\label{m*}
\end{figure}

An examination of the temperature dependence of the amplitude of oscillations in the de Haas-van Alphen (dHvA) effect can give information on the effective mass(es) associated with the extremal portions of the Fermi surface(s) \cite{schoen}. In the iron pnictides, both electron (at the Brillouin zone edges) and hole (at the zone center) Fermi surfaces (pockets) of the essentially two-dimensional electronic structure have been identified both from band structure calculations and from experiment. H. Shishido {\it et al}, a multi-institutional collaboration, \cite{shish} have performed dHvA experiments on the BaFe$_2$(As$_{1-x}$P$_x$)$_2$ series. By extracting the temperature dependent amplitude factor ($X/\sinh X$, $X\propto m^*T/B$) that appears in the Lifshitz-Kosevitch formula \cite{schoen} for the oscillations, they have determined the $x$ dependence of the effective mass on two of the pockets. They argue
these
must be electron pockets at the Brillouin zone edge because the hole scattering rate is probably too large to allow oscillations on the zone-center hole pockets to be detectable \cite{stan}. At low temperatures ($T < 4$ K), $m^*$ appears to diverge as the concentration approaches $x_c\sim 0.33$ in a manner that is not inconsistent with a quantum critical point. It is important to recognize that for essentially all known superconductors (in zero magnetic field) the specific heat shows a mean-field-like jump, which indicates that
the effect of classical critical fluctuations on the mass is minimal.
Thus the observed behavior here may be ascribed to the presence of the magnetic quantum critical point beneath the superconducting dome. As mentioned at the end of section 4, the specific heat, hence $m^*$, should vary as $-\ln (w-w_c)$. So we may expect the experimental effective mass at low temperature to have a form $a - b\ln(x-x_c)$. Figure \ref{m*} shows this fit for one of the pockets for the few points above $x_c$.

The group at Kyoto University has also carried out nuclear magnetic resonance (NMR) experiments on the BaFe$_2$(As$_{1-x}$P$_x$)$_2$ series \cite{NakaiPRB,NakaiPRL}. For the $^{31}$P nucleus, they examined the behavior of nuclear spin-lattice relaxation rate $1/T_1$ and Knight shift $K$ as a function of $x$ and $T$.  $K$ depends on the uniform (${\bf q}=0$) electron susceptibility $\chi({\bf q})$, while $1/T_1$ is determined by the imaginary part of the local (${\bf q}$-averaged) susceptibility. For a Fermi liquid, the Korringa relation $T_1TK^2$ = const = $(\hbar R^2/4\pi k_B)$, where $R$ is the ratio of nuclear to electronic gyromagnetic ratios, holds but it fails in the presence of strong magnetic correlations and/or disorder \cite{bss}. In particular,  AFM correlations can increase $\chi({\bf q}\neq 0)$ thus enhancing $1/T_1$ while not appreciably changing the $\chi({\bf q} = 0)$, which determines $K$ \cite{seb}. In the present experiment, $K$ was found to vary very littl
 e with $x$ and $1/T_1T$ was found to fit well with a Curie-Weiss form $1/T_1T = a + b /(T+\theta)$.
While the fitting parameters $a,b$ vary only slightly with $x$, the ``Curie-Weiss temperature" $\theta$ varies substantially; it is zero at the critical concentration so that the singular part of $1/T_1T$ varies as $1/T$, as is expected for a second-order antiferromagnetic quantum critical point \cite{hvl}.

\begin{figure}[h]
\centering
\includegraphics[totalheight=0.44\textheight, angle=-90, viewport=20 100 400 670,clip]
{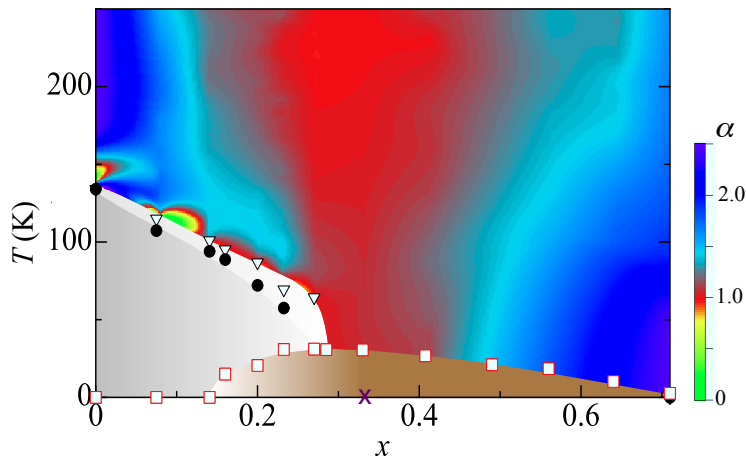}
\caption
{{\bf Phase diagram for BaFe$_2$(As$_{1-x}$P$_x$)$_2$.}
The color shading represents the value of the resistivity exponent in the relation $\rho = \rho_0 + A T^{\alpha}$. The inverted triangles, black dots and squares denote the structural, magnetic (gray region) and superconducting (brown region) transitions, respectively. The purple x marks the quantum critical point at $x_c = 0.33$. Courtesy of Y. Matsuda.}
\label{yuli}
\end{figure}

The Kyoto group has combined their results \cite{NakaiPRL} with earlier resistivity \cite{kasahara} and structural data to produce the informative phase diagram reproduced here in figure \ref{yuli}.

The purple x denotes the quantum critical point at $x_c=0.33$, which is
precisely the value of $x$ at which the measured ``Curie-Weiss temperature"
$\theta$
vanishes. It is also the point where the boundaries of the quantum critical cone, which mark the crossover away from a linear temperature dependence of the resistivity, converge at $T=0$. The fit in figure \ref{m*} shows that it is also the concentration where  a critical divergence of the effective mass occurs, as determined from thermodynamic measurements.


\subsection{1111 compounds}

These are the iron pnictides of composition RFePnO, where R is a rare earth, either non-magnetic (La) or magnetic (Ce) and Pn is a pnictogen.
For LaFeAsO, as the temperature is lowered, there is first, at $T_s$, tetragonal to orthorhombic structural second order transition followed, within 20
degrees, at $T_N$, by a second order magnetic transition to a collinear antiferromagnetic (AFM) state.
In LaFePO, these transitions do not occur: there is no magnetic order down to $T=0$ and there are conflicting reports \cite{cava} about possible superconductivity \cite{kam} below 7 K.
Neither superconductivity nor magnetic order appears in CeFePO.

In the early work \cite{cao}, by the Zhejiang group, on LaFeAs$_{1-x}$P$_x$O, superconductivity ($T_c \sim 10$ K) was found just in the region of the disappearance of magnetism and no information could be obtained on the magnetic quantum critical point, since the upper critical field for the  destruction of superconductivity was rather high ($H_{c2} \sim 27$ T).

The end members of the CeFeAs$_{1-x}$P$_x$O series have the following properties: CeFeAsO \cite{zhao1} has the successive second order transitions mentioned above (structural then magnetic), with $T_s\approx 160$ K and a N{\'e}el temperature $T_N$ of about 135 K. However, CeFePO \cite{brun} has neither magnetic order nor superconductivity. In this compound ($x=1$), the Ce 4$f$ electrons play an important role at low temperature, resulting in heavy fermion properties with a Kondo temperature of about 10 K and ferromagnetic (FM) fluctuations from the 4$f$s. At $x=0$, the Ce $4f$ electrons are antiferromagnetically ordered below 4 K \cite{luo}. This order coexists with that of the Fe $3d$ electrons \cite{jesche}. Across the phase diagram \cite{luo}, as a function of $x$, the Ce $4f$ order changes from AFM to FM in the neighborhood of the Fe $3d$ quantum critical point at $x\approx 0.4$, see below.

\begin{figure}[h]
\centering
\includegraphics[totalheight=0.28\textheight, viewport= 20 10 850 580,clip] {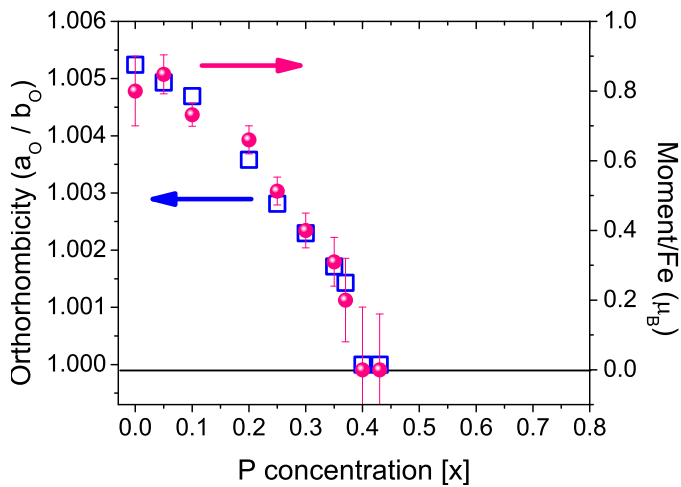}
\caption
{{\bf P-doping dependence of the Fe ordered moments} and the structural orthorhombicity. The proportionality between the two is seen as is the vanishing of both at the the quantum critical point $x_c\approx 0.4$. Adapted with permission from \cite{exp}. Copyright (2010) by the American Physical Society}
\label{CeFig2}
\end{figure}

At Oak Ridge, neutron scattering studies on CeFeAs$_{1-x}$P$_x$O \cite{exp} have shown the behavior of both $T_s$ and $T_N$ as well as the magnitude of the antiferromagnetically ordered moment as functions of P doping for As. The separation of the two transition temperatures, if any, beyond $x = 0.05$ is smaller than the experimental accuracy. $T_s$, $T_N$ and the ordered moment all vanish at $x = 0.4$.
Thus, the behavior is similar to the blue lines of figure \ref{phase}; this compound has a magnetic quantum critical point free from the interference of superconductivity. As is the case for all observations of antiferromagnetism in the pnictides, for the CeFeAs$_{1-x}$P$_x$O system the antiferromagnetic peaks for $0<x<0.4$ are all {\em commensurate}, which supports our use of a local moment picture to describe the magnetism.

The phase diagram ($T$ vs $x$) of CeFeAs$_{1-x}$P$_x$O has been elaborated using transport and thermodynamic measurements by the Zhejiang University group \cite{luo}.
\begin{figure}[h]
\centering
\includegraphics[totalheight=0.3\textheight, viewport=55 80 750 570,clip]
{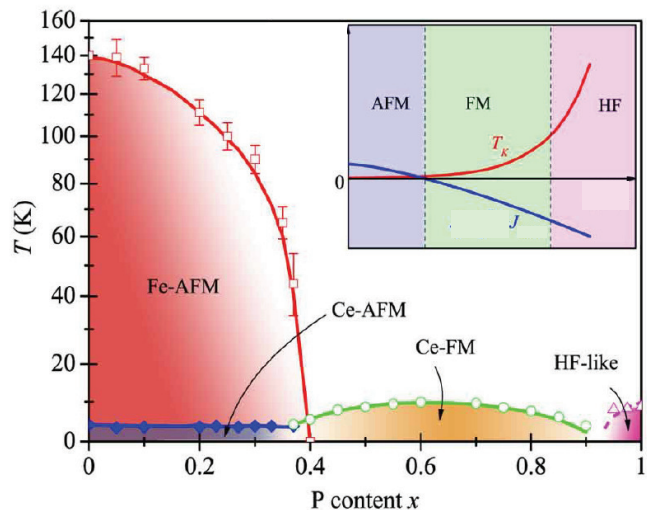}
\caption
{{\bf Phase diagram of CeFeAs$_{1-x}$P$_x$O.}
The inset shows schematically the behavior of the Ce $4f$ electrons as a function of doping. The competition between the Kondo scale ($T_K$) and the exchange coupling (RKKY) scale $J$ is depicted by the red and blue lines. Adapted with permission from \cite{luo}. Copyright (2010) by
the American Physical Society}
\label{phaseAll}
\end{figure}
As is clear from figure \ref{phaseAll}, there is a Fe $3d$ magnetic quantum critical point at $x\approx 0.4$, in agreement with the neutron scattering results \cite{exp}. The very low temperature critical behavior ($T\lesssim 5$ K) at $x\approx x_c$ may be distorted by the onset of order in the Ce $4f$ moments. In contrast to the situation when superconductivity covers the quantum critical point, here the interfering phase belongs to different (Ce $4f$) degrees of freedom than the ones (Fe $3d$) with which we are concerned. Thus there can be ways of separating the effects. The Zhejiang results also confirm the previously observed \cite{brun} Ce $4f$ heavy fermion behavior at $x=1$.

The low temperature behavior of various quantities at the critical concentration will be influenced by the interplay between the onset of Ce magnetism at about 5 K and the Kondo effect with a Kondo temperature of about 10 K.
Nevertheless, it may be of some interest to examine the behavior of the resistivity and susceptibility at higher temperature and $x=x_c$. These quantities have been measured in \cite{luo}. If one tries to describe the resistivity  by a power law $\rho(T)=\rho_0 + A T^{\alpha}$, then $\alpha \approx 1$ only above about $T=125$ K. At lower $T$, $\alpha < 1$,
which in part
reflects the influence of the Kondo effect at low temperatures.
The experimental susceptibility appears to be dominated by the Ce moments and their magnetic transition.

\subsection{Other isoelectronic substitutions}

\begin{figure}[h]
\centering
\includegraphics[totalheight=0.25\textheight, viewport=150 370 750 550,clip]
{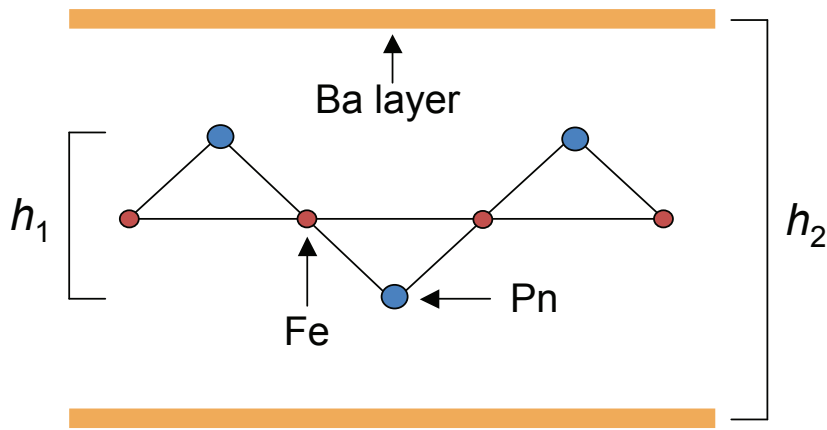}
\caption
{{\bf Schematic of the Ba(FePn)$_2$ (122) crystal structure.}
The Fe-Pn layer thickness is denoted by $h_1$; $h_2$ is one-half the $c$-axis lattice constant. The $a$-axis lattice parameter is $\sqrt{2}$ times the Fe-Fe distance. The pnictogen ions (blue dots) lie alternately above and below the plane of the figure}
\label{122cell}
\end{figure}

The substitution of strontium for barium in the 122 compound is also isoelectronic. Since the  Sr ion is about 17\% smaller than the Ba ion, a reduction of the unit cell volume $V$, similar to that for the P for As doping, should be expected. At the critical concentration $x_c\sim 0.33$ for P-doping, $V$ is about 198 \AA$^3$ \cite{Ba}. That same volume is achieved in Ba$_{1-x}$Sr$_x$Fe$_2$As$_2$ at $x= 0.53$.
However, the behavior is completely different \cite{Ba}. The antiferromagnetic transition temperature $T_N$ {\em increases} as the unit cell volume decreases, rather than decreasing to zero as it does for P-doping. The reason for the different behavior lies in the fact that different forms of chemical pressure affect the internal structure of the unit cell differently. In figure \ref{122cell}, the relevant parameters are shown as $h_1$ and $h_2$. Structural studies \cite{Ba} show that the Fe-pnictogen distance ($h_1/2$)
hardly changes at all under Sr-doping so that all of the volume change comes from a reduction in $h_2 - h_1$, whereas for P-doping, $h_1$ decreases substantially, by about 13\%. Therefore, it is only in the latter case that the Fe-As layer shrinks significantly and of course it is just here that the correlation effects develop.

Another possible isoelectronic substitution is Ru for Fe. In Ba(Fe$_{1-x}$Ru$_x$)$_2$As$_2$, the phase diagram \cite{Sharma,RA,Thaler} for Ru substitution resembles what is depicted in figure \ref{CoDope}. In spite of initial uncertainty about the valence state of Ru, which does not usually occur as $2^+$, there is evidence
that the addition of Ru does not introduce additional $d$-band crossings at the Fermi level and that the total area of hole pockets continues to be essentially the same as that of the electron pockets so that $n=n_h=n_e$. This conclusion can be adduced, for example, from both electronic structure calculation \cite{ZhangSingh} and ARPES experiment \cite{Brouet}. However, the fact that the 4$d$ orbitals of Ru are larger than the 3$d$ orbitals of Fe leads to substantial changes in the important parameters that determine the band structure and the degree of correlation. Thus, the $d-p$ (Ru-As) and $d-d$ overlaps increase causing an increase in bandwidth.
Interestingly,
$n$ appears to be about a factor 2 larger for $x=0.35$ compared to $x=0$ \cite{Brouet}. In addition the effective onsite interaction among the $d$ electrons decreases - both the direct Coulomb repulsion $U$ and the Hund's rule coupling. The latter has been argued to be the main source of correlation effects and the proximity to a Mott transition in the iron pnictides \cite{DMFT}. All this implies that Ru substitution takes the system into the weak correlation regime, so that the origin of the observed disappearance of magnetism cannot be simply ascribed to a bandwidth increase as in the other cases we have discussed, where the relevant electron-electron interactions are unaffected by doping.

\section{Oxychalcogenides}

The chalcogens (Ch) S, Se
combine
with transition metal atoms (Tm) Mn, Fe, Co to form the layered magnetic oxychalcogenide compounds (LaO)$_2$Tm$_2$Ch$_2$O. These compounds are structurally related to the iron pnictides: Tm$_2$Ch$_2$O layers are separated by (LaO)$_2$ layers. The former contains a square lattice of Tm ions and when Tm = Fe, it is similar to the FeAs layer of a pnictide. The Fe valence is again 2+. In the present context, what is interesting about these compounds and analogous ones in which (LaO)$_2$ is replaced by (AF)$_2$ where A = Sr or Ba, is that while structurally similar to the undoped pnictides, they are all insulating with frustrated antiferromagnetism on a checkerboard lattice setting in with $T_N$ of order 100 K. In a combined theoretical and experimental study with collaborators from Los Alamos, Zhejiang, Buffalo, and Rice \cite{oxychal}, we have shown, in agreement with earlier work on (AF)$_2$Fe$_2$Ch$_2$O \cite{kabbour}, that (LaO)$_2$Fe$_2$Se$_2$O is a Mott insulator; this has 
 been confirmed in subsequent studies on this and related compounds \cite{FreeEvans,HuaWu,WangJACS}.
 Therefore, these compounds
have the PE/KE ratio
larger than the value for a Mott transition;
 they are in the Mott insulating state, to the left of the $w$ range shown in figure \ref{phase}.
 The reason that correlation effects dominate in the oxychalcogenides is that the Fe 3$d$ bands are
noticeably
narrower than those in {\it e.g.} LaFeAsO
and FeTe, reflecting a slight (a few percent) expansion of the Fe-square lattice. Thus, while PE is not expected to change much, the KE, measured by the overall bandwidth, is
appreciably
reduced; this drives the material into the Mott state \cite{oxychal}. Thus there is the prospect of accessing a Mott critical point in the oxychalcogenides by increasing KE by, for example, chemical or external pressure.
More generally, the identification of the Mott insulator in materials closely related to
the parent iron arsenides and chalcogenides but with a slightly-expanded Fe-square lattice
supports the description, which involves the importance of electron-electron correlation,
and which forms the basis of the discussion in this review.

\section{Conclusion}

We have summarized the theoretical considerations underlying the prediction for a magnetic quantum critical point in the parent iron pinictides, and the proposal to access it by P doping for As in these materials. The bad-metal nature of the parent iron pnictides and chalcogenides places them in the intermediate-interaction region, thus in proximity to a Mott transition. In this regime, the spin spectrum comes primarily from electronic states away from the Fermi surfaces; we model this by a $J_1-J_2$ spin Hamiltonian. At the same time, the
electronic spectral weight in the vicinity of the Fermi surfaces is tuned by the strength of interactions, and this tuning gives rise to a magnetic quantum critical point.

Isoelectronic doping such as P for As in the parent iron pnictides provides an ideal setting to access this quantum critical point. We have gathered the experimental evidence that has since been accumulated for such a quantum critical point. In Ce-1111 materials, P doping for As does not induce superconductivity in the measured temperature range (above 2K). Neutron scattering measurements of the commensurate antiferromagnetic order parameter, combined with transport measurements of the temperature-doping phase diagram, make a compelling case for the existence of a magnetic quantum critical point at a P doping about $0.4$. In 122 materials, superconductivity is observed in the intermediate range of P dopings. Still, the combination of transport, NMR, and quantum oscillation measurements provide a strong evidence for a quantum critical point in the temperature-doping phase diagram, at a P doping of about $0.33$,  that is however masked by the superconducting dome.

In carrier-doped iron pnictides, evidence for quantum criticality has been advanced
 \cite{jhchu,gooch} although the case appears not yet as strong as for the isoelectronically-doped parent pnictides, which we have discussed here. Moreover, the carrier-doped materials are
 more disordered than the isoelectronically-doped ones, so that the latter are more
 definitively studied by various probes.

 The existence of a quantum critical point in the overall phase diagram of these materials has important consequences.
 The  associated
 quantum criticality serves as a
likely source of non-Fermi liquid behavior, such as the anomalous temperature dependence of electrical resistivity,
that has been observed in several different parts of the iron pnictide phase diagrams.
It  provides a proper characterization of the magnetic excitations in the iron pnictides and chalcogenides, and this is important for understanding the interplay among magnetism, localization, and superconductivity. At the same time, it unveils a new setting to advance our general understanding of quantum critical phenomena.

Finally, the isoelectronically-tuned magnetic quantum criticality represents
a concrete prediction of the strong-coupling approach that
has been verified by subsequent experiments; this, together with the demonstration of a Mott localization induced by a slight expansion of the Fe-square lattice that has also been summarized here, underscores the viability of this approach to the microscopic physics of iron-based superconductors.

 We thank J. Dai, P. Goswami, P. Nikolic, R. Yu and J-X Zhu for collaborations and discussions. This work has been partially supported by the NSF Grant No. DMR-1006985 and the Robert A. Welch Foundation Grant No. C-1411. We also acknowledge the hospitality of Aspen Center of Physics, where part of the work was carried out.


\begin{thebibliography}{99}
\bibitem{foot1}According to band structure calculations, all 5 iron $d$-orbitals are present at the Fermi surface. Therefore the exchange couplings are matrices in the orbital space and a Hund's rule term should be included in ${\cal H}_{mag}$, see
\cite{sieaprl,njp}. Here, we use a one-orbital Hamiltonian to illustrate the physics.
\bibitem{foot2}From (1), $\tau \propto \xi^z$. Similarly, in the action of (\ref{gl}), the ratio of the smallest powers of  $q$ and $\omega$ is $z$.
\end{thebibliography}
\end{document}